\documentstyle[aps,prl,multicol,epsfig,fancyheadings]{revtex}
\begin{document}

\draft
\twocolumn[\hsize\textwidth\columnwidth\hsize\csname@twocolumnfalse\endcsname
\title{
The Quantum--Mechanical Position Operator in Extended Systems
}
\author{Raffaele Resta}

\address{INFM--Dipartimento di Fisica Teorica, Universit\`a di Trieste, Strada
Costiera 11, 34014 Trieste, Italy \\ Department of Physics, The Catholic
University of America, Washington, D.C. 20064}

\date{September 15, 1997} 
\maketitle

\begin{abstract} The position operator (defined within the Schr\"odinger
representation in the standard way) becomes meaningless when periodic
boundary conditions are adopted for the wavefunction, as usual in condensed
matter physics. We show how to define the position expectation value by means
of a simple many--body operator acting on the wavefunction of the extended
system. The relationships of the present findings to the Berry--phase theory
of polarization are discussed.  \end{abstract}

\bigskip\bigskip

]
\narrowtext

The position operator within the Schr\"odinger representation acts
multiplying the wavefunction by the space coordinate. This is trivial, but
only applies to the bound eigenstates of a finite system, which belong to the
class of square--integrable wavefunctions. This is not the way how condensed
matter theory works: almost invariably, one considers a large system within
periodic boundary conditions (PBC), and the position operator (defined as
usual) becomes then meaningless. For the sake of simplicity, most of this
Letter will deal with the one--dimensional case. The Hilbert space of the
single--particle wavefunctions is defined by the condition $\psi(x+L) =
\psi(x)$, where $L$ is the imposed periodicity, chosen to be large with
respect to atomic dimensions. An operator maps any vector of the given
space into another vector belonging to the same space: the multiplicative
position operator $x$ is {\it not} a legitimate operator when PBC are adopted
for the state vectors, since $x \,\psi(x)$ is not a periodic function
whenever $\psi(x)$ is such. Of course, any periodic function of $x$ is a
legitimate multiplicative operator: this is the case {\it e.g.} of the
nuclear potential acting on the electrons.  Since the position operator is
ill defined, so is its expectation value, whose observable effects in
condensed matter are related to macroscopic polarization. For the crystalline
case, the long--standing problem of dielectric polarization has been solved a
few years ago~\cite{modern,rap_a12,Ortiz94}: polarization is a manifestation
of the Berry phase~\cite{Berry,rap_a17}, {\it i.e.} is an observable which
cannot be cast as the expectation value of {\it any} operator, being instead
a gauge--invariant phase of the wavefunction. Here we find a different, and
more fundamental, solution: we arrive indeed at defining the expectation
value of the position in an extended quantum system within PBC, where the
operator entering this definition is simple but rather peculiar. Amongst the
most relevant features, the expectation value is defined modulo $L$, and the
operator is no longer one--body: it acts as a genuine many--body operator on
the periodic wavefunction of $N$ electrons. In the crystalline case, the
present result can be related to a discretized Berry phase, and sheds new
light into the physical meaning of the latter.

We study a system of $N$ electrons in a segment of length $L$, and eventually
the thermodynamic limit is taken: $L \rightarrow \infty$, $N \rightarrow
\infty$, and $N/L = n_0$ constant. At any finite $L$ the ground eigenfunction
obeys PBC in each electronic variable separately: \begin{equation} \Psi_0(x_1,
\dots, x_i, \dots, x_N) = \Psi_0(x_1, \dots, x_i\! + \! L, \dots, x_N) . 
\label{perio} \end{equation} We assume the ground state nondegenerate, and we
deal with insulating systems only: this means that the gap between the ground
eigenvalue and the excited ones remains finite for $L \rightarrow \infty$. 
Since the spin variable is irrelevant to this problem, we omit it altogether
and we deal with a system of spinless electrons. Our major goal is defining
the expectation value of the electronic position $\langle X \rangle$, and to
prove that our definition provides in the thermodynamic limit the physical
macroscopic polarization of the sample.  

Before attacking the main problem, let us discuss the much simpler case where
PBC are {\it not} chosen, and the $N$-particle wavefunction (called $\Phi_0$
in this case) goes to zero exponentially outside a bounded region of space. We
may safely use the operator $\hat{X} = \sum_{i=1}^N x_i$, and define the
position expectation value as usual: \begin{equation} \langle X \rangle =
\langle \Phi_0 | \hat{X} | \Phi_0 \rangle = \int \!dx \; x \, n(x),
\label{trivial} \end{equation} where $n(x)$ is the one--particle density. The
value of $\langle X \rangle$ scales with the system size, and the quantity of
interest is indeed the dipole per unit length, which coincides with
macroscopic polarization.  The operator $\hat{X}$, being the sum of identical
operators acting on each electronic coordinate separately, is by definition a
one--body operator. The expectation value of this same operator cannot be
evaluated if the wavefunction obeys PBC: in fact $\hat{X}$ does not
commute with a translation by $L$, and therefore is not a legitimate operator
in the Hilbert space defined by Eq.~(\ref{perio}).

We are now ready to state our main result, demonstrated in the following. 
When PBC are adopted, the position expectation value
must be defined through: \begin{equation} \langle X \rangle = \frac{L}{2\pi}
\mbox{Im log }  \langle \Psi_0 | {\rm e}^{i\frac{2\pi}{L} \hat{X}} | \Psi_0
\rangle . \label{main} \end{equation} The expectation value $\langle X
\rangle$ is thus defined only modulo $L$, hardly a surprising finding since
the wavefunction is periodical.  The operator occurring in Eq.~(\ref{main}) is
a legitimate one in the Hilbert space of periodic wavefunctions; as a tradeoff
it is no longer the sum of operators acting on each electronic coordinate
separately, and is therefore a genuine many-body operator.  At variance with
Eq.~(\ref{trivial}), the expectation value $\langle X \rangle$ within PBC
{\it cannot} be expressed in terms of the one--particle
density, not even in terms of the one--particle reduced density matrix. The
$N$-particle wavefunction is explicitly needed.

What remains to be done is to prove that our definition of $\langle X
\rangle$ provides the relevant physical observable in the thermodynamic
limit. Notice that $L \rightarrow \infty$ is a tricky limit, since the
exponential operator in Eq.~(\ref{main}) goes formally to the identity, but
the size of the system and the number of electrons in the wavefunction
increase with $L$~\cite{nota}.  We will show that the electronic polarization
(dipole per unit length) is: \begin{equation} P_{\rm el} =  \lim_{L
\rightarrow \infty} \frac{e}{2\pi} \mbox{Im log }  \langle \Psi_0 | {\rm
e}^{i\frac{2\pi}{L} \hat{X}} | \Psi_0 \rangle , \label{limit} \end{equation}
where $e$ is the electron charge.  It is expedient to introduce the family of
Hamiltonians: \begin{equation} \hat{H}(\alpha) = \frac{1}{2m}\sum_{i=1}^N
(p_i - \hbar \alpha)^2 + \hat{V} , \label{family} \end{equation} where
$\alpha$ is a real constant, and $\hat{V}$ is the many--body potential. A
parametric Hamiltonian of this kind was first introduced by W. Kohn many
years ago~\cite{Kohn64}, and subsequently used by different
authors~\cite{Niu84,Ortiz94}. The ground eigenstate of $\hat{H}(0)$ is
precisely $| \Psi_0 \rangle$; more generally, the state vector ${\rm
e}^{i\alpha \hat{X}} | \Psi_0 \rangle$ fulfills the equation \begin{equation}
\hat{H}(\alpha) \; {\rm e}^{i\alpha \hat{X}} |  \Psi_0 \rangle = E_0 \; {\rm
e}^{i\alpha \hat{X}} | \Psi_0 \rangle ,\label{sch} \end{equation} with an
$\alpha$-independent $E_0$.  This does not warrant that it is an eigenstate
since PBC, Eq.~(\ref{perio}), are {\it not} fulfilled, except in the special
cases where $\alpha$ is a multiple of $2\pi / L$. Since by hypothesis $E_0$
is nondegenerate,  ${\rm e}^{i 2 \pi \hat{X}/ L} | \Psi_0 \rangle$ is the
ground eigenstate of $\hat{H}(2\pi / L)$: we may then use perturbation theory
to expand it to leading order in $1/L$ in terms of the eigenstates  $| \Psi_j
\rangle$ of $\hat{H}(0)$. However, the standard formulas perform an arbitrary
choice for the phase of the perturbed eigenstate; in the most general case we
write instead: \begin{equation} {\rm e}^{i\frac{2 \pi }{L} \hat{X}} | \Psi_0
\rangle \simeq {\rm e}^{i\gamma_L} \Bigl( | \Psi_0 \rangle - \frac{2 \pi
\hbar}{mL}\sum_{j \neq 0} | \Psi_j \rangle \frac{ \langle \Psi_j | \hat{P} |
\Psi_0 \rangle }{ E_0 - E_j} \Bigr) , \label{pert} \end{equation} where
$\hat{P} =  \sum_{i=1}^N p_i$ is the momentum operator.  It is important to
realize that the perturbative expansion is a good approximation whenever $L$
is much larger than a typical atomic dimension, while the number of electrons
in the wavefunction and the system size are irrelevant.

Replacement of Eq.~(\ref{pert}) into Eq.~(\ref{main}) shows that the phase
$\gamma_L$ is a most fondamental quantity, since: \begin{equation} \langle X
\rangle \simeq \frac{L \gamma_L}{2 \pi} , \label{gamma} \end{equation} but we
have not yet related it to any observable. In order to prove that $e \langle X
\rangle / L$ for large $L$ is indeed the electronic polarization---as
anticipated in Eq.~(\ref{limit})---it will be enough to show that its time
derivative coincides with the adiabatic electrical current flowing through the
system whenever the Hamiltonian contains a slowly varying time--dependent
term. We start from \begin{equation} \frac{d}{dt} \langle X \rangle =
\frac{L}{2\pi} \mbox{Im} \Bigl( \frac{\langle \dot{\Psi}_0 | {\rm
e}^{i\frac{2\pi}{L} \hat{X}} | \Psi_0 \rangle} {\langle \Psi_0 | {\rm
e}^{i\frac{2\pi}{L} \hat{X}} | \Psi_0 \rangle} + \frac{\langle \Psi_0 | {\rm
e}^{i\frac{2\pi}{L} \hat{X}} | \dot{\Psi_0} \rangle} {\langle \Psi_0 | {\rm
e}^{i\frac{2\pi}{L} \hat{X}} | \Psi_0 \rangle} \Bigr), \label{deriv}
\end{equation} where $|\dot{\Psi}_0 \rangle$ is the time derivative of the
instantaneous adiabatic eigenstate. Substituting now Eq.~(\ref{pert}) in
Eq.~(\ref{deriv}) the phase factor cancels out; to lowest order in $1/L$ we
get \begin{equation} \frac{e}{L}\frac{d}{dt} \langle X \rangle \simeq \frac{i
e \hbar}{mL} \sum_{j \neq 0} \langle \dot{\Psi}_0 | \Psi_j \rangle \frac{
\langle \Psi_j | \hat{P} | \Psi_0 \rangle }{ E_0 - E_j} + \; \rm{c.c.} \; ,
\label{repl} \end{equation} where c.c. indicates the complex conjugate.  In
Eq.~(\ref{repl}) the $j\!=\!0$ term is omitted from the sum, since
$|\dot{\Psi}_0 \rangle$ can be taken as orthogonal to $| \Psi_0 \rangle$ with
no loss of generality; furthemore we have exploited time--reversal symmetry,
owing to which all the adiabatic instantaneous eigenstates $\Psi_j$ can be
taken as real.

This concludes our proof. In fact the right--hand member of Eq.~(\ref{repl})
is the electronic current flowing through the system when the potential
$\hat{V}$ is adiabatically varied, a well known expression due to Thouless:
Eq.~(2.5) in Ref.~\onlinecite{Thouless83}. The rest of this Letter is devoted
to an analysis of our major result, Eqs.~(\ref{main}) and (\ref{limit}), and
of its relationship to previous work.

The special case of $N\!=\!1$ corresponds to a lone quantum electron diluted
in a large sample. The position expectation value, Eq.~(\ref{main}), can then
be expressed in terms of the periodic density as: \begin{equation} \langle X
\rangle = \frac{L}{2\pi} \mbox{Im log }  \int_0^L \!\! dx \;  {\rm
e}^{i\frac{2\pi}{L} x} n(x) . \label{one} \end{equation} A similar expression
has been previously used by a few authors\cite{heuris} in order to
heuristically follow the adiabatic time evolution of a single quantum
particle in a disordered condensed system within PBC. The case $N\!>\!1$ is
{\it qualitatively} different, in that---as stressed above---the operator
used in Eq.~(\ref{main}) to define $\langle X \rangle$ is a genuine
many--body one.  This is particularly remarkable in view of the fact that the
physical observable is an integrated current: the current is a typical
one--body operator, as in fact is the operator $\hat{P}$ in the right--hand
member of Eq.~(\ref{repl}).  The case of independent electrons is also worth
commenting. In this special case the $N$-particle wavefunction is uniquely
determined by the one--body reduced density matrix $\rho(x,x')$ (which is the
projector over the set of the occupied one--particle orbitals): therefore the
expectation value $\langle X \rangle$ is uniquely determined by $\rho$.

For the crystalline case, macroscopic polarization is presently understood as
a manifestation of the Berry phase\cite{Berry}, both for independent
electrons\cite{modern,rap_a12} and for correlated
electrons\cite{Ortiz94,rap_a17}. The definition of Eq.~(\ref{limit}) reduces
to the well established ones in the crystalline case: for finite $L$ the
present findings can be shown to be equivalent to a discretization of the line
integral defining the Berry phase.  In this Letter we provide an explicit
proof for the independent--electrons case only: the correlated case is not
much different.

Suppose we have a crystalline system of lattice constant $a$, where we impose
PBC over $M$ linear cells: there are then $M$ equally spaced Bloch vectors in
the reciprocal cell  $[0,  2 \pi / a)$: \begin{equation} q_s = \frac{2 \pi}{M
a} s, \quad s=0,1,\dots, M\! -\!1 .  \label{points} \end{equation} The size
of the periodically repeated system is $L = M a$.  The one--body orbitals can
be chosen to have the Bloch form: \begin{equation} \psi_{q_s,m}(x+ \tau) =
{\rm e}^{i q_s \tau} \psi_{q_s,m}(x) , \label{bloch} \end{equation} where
$\tau = la$ is a lattice translation, and $m$ is a band index. There are
$N/M$ occupied bands in the Slater determinant wavefunction, which we write
as \begin{equation} | \, \Psi_0 \rangle = {\sf A} \prod_{m=1}^{N/M} \;
\prod_{s=0}^{M-1} \psi_{q_s,m} , \label{det} \end{equation}  where ${\sf A}$
is the antisymmetrizer. It is now expedient to define a new set of Bloch
orbitals: \begin{equation} \tilde{\psi}_{q_s,m}(x) = {\rm e}^{- i
\frac{2\pi}{L} x} \psi_{q_s,m}(x) .  \label{tilde} \end{equation} We then
recast the expectation value of Eq.~(\ref{main}), after a double change of
sign, as: \begin{equation} \langle X \rangle = - \frac{L}{2\pi} \mbox{Im log
}  \langle \Psi_0 | \tilde{\Psi}_0 \rangle , \label{main2} \end{equation}
where $| \tilde{\Psi}_0 \rangle$ is the Slater determinant of the
$\tilde{\psi}$'s. According to a well known theorem, the overlap amongst two
determinants is equal to the determinant of the overlap matrix amongst the
orbitals: \begin{equation} \langle X \rangle = - \frac{L}{2\pi} \mbox{Im log
det}\; S , \label{main3} \end{equation} where \begin{equation} S_{sm,s'm'} =
\int_0^L\!\!dx \; \psi^*_{q_s,m}(x) {\rm e}^{- i \frac{2\pi}{L} x}
\psi_{q_{s'},m'}(x) \label{overlap}. \end{equation} Owing to the
orthogonality properties of the Bloch functions, the overlap matrix elements
vanish except when $q_{s'} = q_s + 2\pi /L$, that is $s' = s\!+\!1$.  The
$N\!\times\!N$ determinant can then be factorized into $M$ small
determinants: \begin{equation} \mbox{det} \; S = \prod_{s=0}^{M-1} \mbox{det}
\; S(q_s,q_{s+1}) , \end{equation} where---in order to  make contact with
previous literature\cite{rap_a12}---for the small overlap matrix we use the
notation \begin{equation} S_{m,m'}(q_s,q_{s+1}) = \int_0^L\!\!dx \;
\psi^*_{q_s,m}(x) {\rm e}^{- i \frac{2\pi}{L} x} \psi_{q_{s+1},m'}(x)
\label{overlap2}, \end{equation} and $\psi_{q_{M},m}(x) \equiv
\psi_{q_0,m}(x)$ is implicitly understood (so--called periodic gauge). 
Replacing Eq.~(\ref{overlap2}) into Eq.~(\ref{limit}) we get \begin{equation}
P_{\rm el} = -  \frac{e}{2\pi} \lim_{L \rightarrow \infty} \mbox{Im log}
\prod_{s=0}^{M-1} \mbox{det} \; S(q_s,q_{s+1}) , \label{limit2}
\end{equation} which concludes our equivalence proof.  Eq.~(\ref{limit2})
coincides in fact with the well known expression of the modern theory of
polarization~\cite{modern,rap_a12}, obtained by King--Smith and Vanderbilt by
defining a (continuum) Berry phase as a line integral, and then discretizing
it.  

The discretization of the Berry phase was originally introduced for purely
computational purposes, and is in fact routinely used in first--principles
calculations~\cite{rap_a12}. The alternate path followed here to arrive at
the same result shows that the discretization has instead a very basic
meaning of its own.  Macroscopic polarization can be cast as the
thermodynamic limit of an expression involving the expectation value of a
relatively simple and physically meaningful many-body operator as in
Eq.~(\ref{limit}). This operator ``extracts'' the Berry phase from the {\it
square modulus} of the many--body wavefunction, which embeds the relevant
information about the {\it relative phases} of the one--particle orbitals.

Several previous findings about macroscopic
polarization\cite{modern,rap_a12,Ortiz94} apply to the present formulation as
well: we report here a few of them for the sake of completeness. One is not
interested in defining an ``absolute'' polarization: the measured bulk
quantity is always the difference $\Delta P$ between two states of the given
solid, connected by an adiabiatic transformation of the Hamiltonian:
\begin{equation} \Delta P = \Delta P_{\rm nucl} + \Delta P_{\rm el} =
\int_0^{\Delta t} \!\! dt \; J(t), \label{current} \end{equation} where
$J(t)$ is the total (nuclear + electronic) current flowing throught the
sample while the potential $\hat{V}$ is adiabatically varied. Notice that in
the adiabatic limit $\Delta t$ goes to infinity and $J(t)$ goes to zero. The
quantity of interest  $\Delta P$ can be evaluated as a two--point formula,
using the initial and final states only: for the electronic term, one
evaluates Eq.~(\ref{limit}) with both the final and the initial
wavefunctions, and takes the difference. The result is only defined modulo
$e$; a similar indeterminacy applies to the nuclear term $\Delta P_{\rm
nucl}$.  Nothing can be done about this ambiguity of the two--point formula,
which ultimately stems from Thouless's quantization of particle
transport\cite{Thouless83,Niu84,Pendry84}. There is of course no
indeterminacy if one trades away the two--point formula and performs instead
the time integral in Eq.~(\ref{current}), using for the electronic current
the right--hand term of Eq.~(\ref{repl}).

Generalization of Eqs.~(\ref{main}) and (\ref{limit}) to the
three--dimensional case requires some care. Since $\langle X \rangle$ is
extensive, within a na\"{\i}f approach the $x$-component of $\Delta {\bf
P}_{\rm el}$ would be defined only modulo $e/L^2$, which becomes vanishingly
small in the thermodynamic limit. Fortunately, the problem is less serious
than this, and the drawback is easily eliminated by adapting to the present
formulation a major finding from Ref.~\onlinecite{Ortiz94}. Suppose the
system is crystalline with a simple cubic lattice of constant $a$. 
Then---upon exploiting the lattice periodicity of the one--particle
density---it can be shown that the two--point formula provides each component
of $\Delta {\bf P}_{\rm el}$ with an indeterminacy of $e/a^2$, which is no
serious drawback. Such indeterminacy has nothing to do with electron
correlation, and not even with quantum mechanics: a similar indeterminacy is
also present in the classical nuclear term $\Delta {\bf P}_{\rm nucl}$
whenever this term is evaluated as a two--point formula\cite{nota2}. If the
system is noncrystalline, then a large ``supercell'' is needed to reproduce
the disorder, and only small polarization differences are accessible via the
two--point formula. Again, this looks like a fundamental consequence of
Thouless's quantization of particle transport\cite{Thouless83,Niu84,Pendry84}.

The modern viewpoint about macroscopic
polarization\cite{modern,rap_a12,Ortiz94} has even spawned a critical
rethinking of density--functional theory in extended systems. The debate
started in 1995 with a paper by Gonze, Ghosez, and Godby~\cite{Gonze95}, and
continues these days\cite{debate}. Although the subject is clearly outside
the scope of this Letter, I point out that the present main
achievement---namely, defining the polarization of a many-electron system by
means of an expectation value---could possibly help in further clarifying the
matter.

Discussions with M. Bernasconi and E. Yaschenko are gratefully acknowledged.
Work partly supported by by the Office of Naval Research, through grant
N00014-96-1-0689.


\begin{references}
\bibitem{nota2} The indeterminacy of the nuclear term is $e/a^2$ times a
nuclear charge.
\end{references}
\end{document}